\documentclass[ps,pra,showpacs,superscriptaddress,twocolumn]{revtex4-2}

\usepackage[utf8]{inputenc}
\usepackage{amsmath}
\usepackage{amssymb}
\usepackage{color}
\usepackage{verbatim}
\usepackage{array}
\usepackage{graphicx}
\usepackage{bm}
\usepackage[ragged]{sidecap}
\usepackage{dsfont}
\usepackage{ulem}
\usepackage{mathrsfs}
\usepackage{gensymb}
\usepackage{siunitx}
\usepackage{url}
\usepackage{dcolumn}
\usepackage{soul}
\usepackage[commandnameprefix=ifneeded,final]{changes}
\newcolumntype{d}{D{.}{.}{-1}}
\newcolumntype{f}[1]{D{.}{.}{#1}}

\usepackage{hyperref}
\hypersetup{
    colorlinks,
    citecolor=blue,
    filecolor=orange,
    linkcolor=cyan,
    urlcolor=green}
\usepackage{orcidlink}

\newcommand{\ie}{{\textit{i.e., }}}

\begin{document}

%\preprint{APS/123-QED}

\title{A Bayesian Approach for Strong Field QED Tests with He-like Ions}

\author{C\'esar Godinho\orcidlink{0000-0003-3195-0394}}
\email{c.godinho@campus.fct.unl.pt}
\affiliation{Laboratory of Instrumentation, Biomedical Engineering and Radiation Physics (LIBPhys-UNL),Department of Physics, NOVA School of Science and Technology, NOVA University Lisbon, 2829-516 Caparica, Portugal}

\author{Jorge Machado\orcidlink{0000-0002-0383-4882}}
\email{jfd.machado@fct.unl.pt}
\affiliation{Laboratory of Instrumentation, Biomedical Engineering and Radiation Physics (LIBPhys-UNL),Department of Physics, NOVA School of Science and Technology, NOVA University Lisbon, 2829-516 Caparica, Portugal}

\author{Nancy Paul\orcidlink{0000-0003-4469-780X}} 
\email{nancy.paul@lkb.upmc.fr}
\affiliation{Laboratoire Kastler Brossel, Sorbonne Universit\'e, CNRS, ENS-PSL Research University, Coll\`ege de France, Case\ 74;\ 4, place Jussieu, F-75005 Paris, France}

\author{Mauro Guerra\orcidlink{0000-0001-6286-4048}} 
\email{mguerra@fct.unl.pt}
\affiliation{Laboratory of Instrumentation, Biomedical Engineering and Radiation Physics (LIBPhys-UNL),Department of Physics, NOVA School of Science and Technology, NOVA University Lisbon, 2829-516 Caparica, Portugal}

\author{Paul Indelicato\orcidlink{0000-0003-4668-8958}} 
\email{paul.indelicato@lkb.upmc.fr}
\affiliation{Laboratoire Kastler Brossel, Sorbonne Universit\'e, CNRS, ENS-PSL Research University, Coll\`ege de France, Case\ 74;\ 4, place Jussieu, F-75005 Paris, France}
\homepage{http://www.lkb.upmc.fr/metrologysimplesystems/project/paul-indelicato/}

\author{Martino Trassinelli\orcidlink{0000-0003-4414-1801}} 
\email{martino.trassinelli@insp.jussieu.fr}
\affiliation{Institut des NanoSciences de Paris, CNRS, Sorbonne Université, F-75005 Paris, France}

\date{\today}

\begin{abstract}
Detailed comparisons between theory and experiment for quantum electrodynamics (QED) effects in He-like ions have been performed in the literature to search for hints of new physics. 
Different frequentist statistical analyses of the existing atomic transition energy data have shown contradictory conclusions as to the presence of possible deviations from the theory
predictions.
We present here an approach using Bayesian statistics \deleted{which allows} to assign quantitative probabilities to the different deviation models from theory 
for He-like ions for $Z = 5$ to \num{92}. 
Potential deviations
beyond the standard model or higher-order QED effects are modeled with $f(Z) \propto Z^k$ functions. Considering the currently available data, no significant difference between theory and experiment is found, and we show that recent experiments have reduced the possible deviations previously observed in the literature. 
Using past measurements and a weighted average on the different deviation models, we estimate the accuracy required for future measurements to investigate possible \replaced{anomalies}{divergences}.
\end{abstract}

\maketitle

\section{Introduction}

Quantum electrodynamics (QED) is one of the foundations of contemporary physics, and a complete understanding of this theory is key for searches for new physics with atoms and molecules \cite{sbdk2018}. The advent of high precision physics necessitates a correspondingly accurate theory for predicting the quantum structure of atoms that results from interactions with the field fluctuations of the vacuum\added{, manifest as, for example, self-energy and vacuum polarization effects. 
For light atoms, this is normally carried out by starting from the non-relativistic solutions and considering relativistic and QED effects as a perturbation.
This results in an expansion as a function of the nuclear binding strength parameter $Z \alpha$, where $Z$ is the nuclear charge number and $\alpha$ is the fine-structure constant.
For heavy atoms and ions, this perturbative approach cannot be carried out, as the expansion parameter $Z \alpha$ gets close to unity. 
Non-perturbative techniques, with respect to $Z \alpha$, must be applied in these cases,leaving only an expansion with respect to the electromagnetic coupling, proportional to $\alpha \approx 1/137$.}

%\added{An additional difficulty for the theoretical predictions is added when more than one bound electron is present.}
As solving the QED many-body problem is exceedingly complex, notably when going to strong Coulomb fields, experiments with atoms with one or two electrons play an important key role in testing the implementation of the theory. 
Few-body highly charged ions (HCIs) are a privileged testing ground because they allow the interplay between QED and electron--electron (e-e) correlation effects to be explored at the same time.
In the high-$Z$ region, the uncertainty of the nuclear size contribution to the energy values of the transitions adds an additional ingredient to be disentangled.
 
Decades of effort with diverse experimental methods have aimed at precision spectroscopy of He-like ions. 
Most of the available data for He-like ions lies between $12 \leq Z \leq 60$, with few points for the highest-$Z$ elements, and few points with accuracy matching those of lower Z.
Studies in the low-$Z$ regime are generally characterized by high precision and obtained by laser spectroscopy.
Measurements on heavy HCIs are performed at large-scale accelerators and storage rings.
\added{Because of the huge binding energy of the electrons in such systems, up to about 100~keV, the ionization of such heavy ions can be obtained only by high-energy collisions.
Beams of heavy few-electron ions are obtained in reasonable quantities ($10^8$ ions per bunch)} by stripping bound electrons in thin foils at kinetic energies of few hundred MeV/u.
\added{The ions are subsequently accumulated in storage rings where their characteristic radiative emission is measured.
Due to the relatively high velocity of the ion beams, more than 10\% of the speed of light even after applying deceleration methods, such measurements are characterized by high uncertainties} coming from the relativistic Doppler correction, in addition to relatively poor energy resolutions of standard solid-state detectors, commonly used for this energy range.

Because the different contributions to the transition energies vary with different powers of the atomic number $Z$, systematic measurements across a wide range of $Z$ are needed to reliably constrain the theory and disentangle one-electron QED effects from e-e and nuclear ones (see for example Refs.~\cite{bab2015}, \cite{ind2019} for recent reviews). 
\added{Moreover, it has recently been demonstrated that non-perturbative and perturbative approaches, in $Z \alpha$, provide contradictory predictions for higher-order self-energy contributions \cite{Yerokhin2024}.
A systematic study with respect to $Z$ is thus needed to further investigate this issue.}

Numerous authors have performed analyses of the existing data trying to evaluate potential systematic discrepancies as a function of $Z$ that would point to missing physics \cite{cak2009,ckgh2012,cpgh2014,bab2015,ind2019}. Such studies have historically been performed within frequentist statistical methods, where an ansatz is made about the model to describe the (dis)agreement between theory and experiments as a function of $Z$, and a quantitative analysis is performed based on the $\chi^2$ goodness-of-fit test \cite{ckgh2012, cpgh2014, bab2015, mssa2018,ind2019}, \ie relying only on the minimal value of $\chi^2$ functions (or the maximal value of likelihood functions), implicitly assuming certain regularity, like mono-modality, of such functions. The vigorous debate around this question attests to the importance of this investigation.
However, it might be dangerous to draw conclusions with data from a limited range of $Z$ and with relatively few precision measurements.
A comparison between different analyses is further complicated by the fact that different authors use different data sets and different criteria to evaluate model significance. 
Moreover, the use of statistical criteria can be a problem when there is no propensity for one particular model.

In this article, we propose an approach to remedy the aforementioned problems based on quantitative Bayesian statistical methods.
We have considered all data from the statistical analysis of $n=2 \to 1$ transitions from the recent compilation \cite{ind2019} with the inclusion of very recent experiments~\cite{mpsd2023,Pfafflein2024} and measurements that were missing in previous analyses~\cite{hbwg2016}. In addition, measurements of $n=2 \to 2$ intrashell transitions, where the scaling laws of QED contributions are different, have also been analysed, including the recent improved measurement on He-like uranium \cite{lbds2024}. 

\section{Methods}

\begin{figure*}[t]
\includegraphics[width=\textwidth]{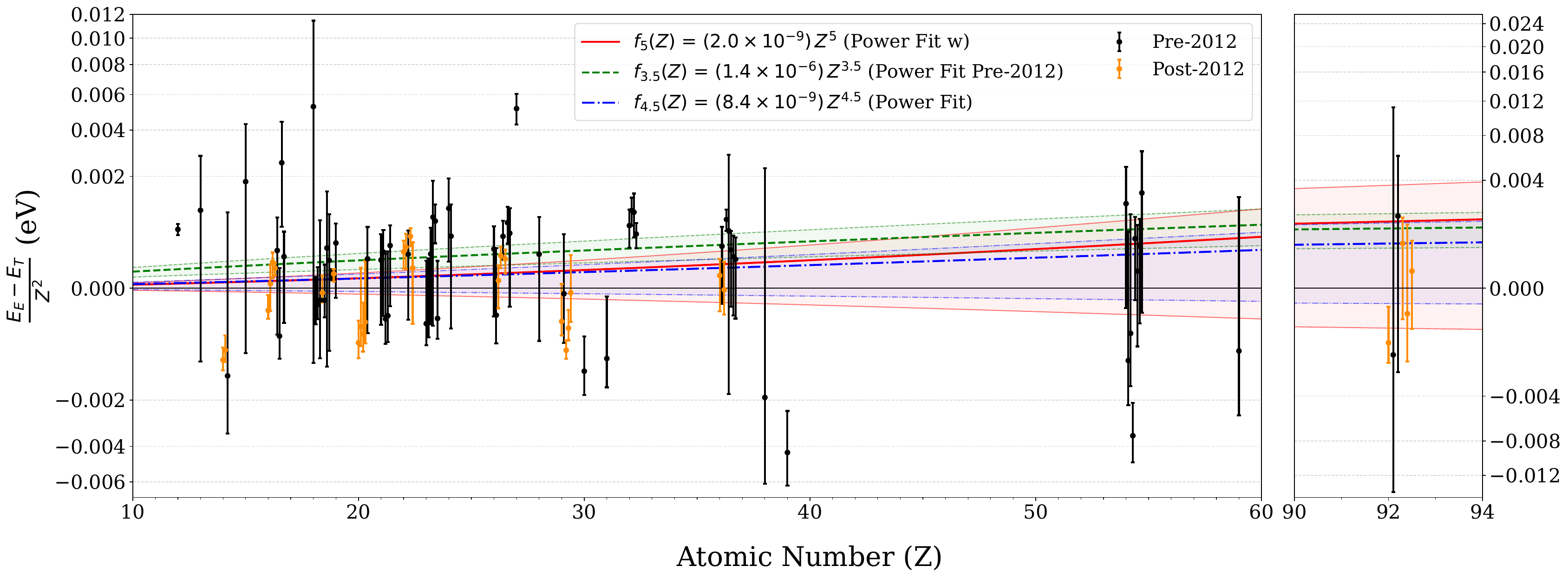}
\caption{\label{fig:fit}The most likely (highest evidence) fits for the measurements from Refs. \cite{aamp1988,abzp1974,asgl2012,bab2015,bbgh1989,bcid1990,biss1989,bitg1984,bmic1983,btmd1984,cphs2000,dbf1984,dlp1978,eal1995,esbr2015,itbl1986,kmmu2014,ldhs1994,mbvk1992,mssa2018,pckg2014,rbes2013,rgtm1995,rrag2014,sbbt1982,tbcc2008,tgbb2009,wbbc2000,wbdb1996,mpsd2023,hbwg2016, pgsz2024}. The red line represents the most likely fit considering the subset of data consisting of w lines, the green line represents the most likely fit considering all measured lines in the literature until the year of 2012 and the blue line the fit considering all available experimental data. The respective colored shadow regions display a 99\% confidence interval for each fit. Coincident measurements on the x-axis are shifted to improve readability. The y-axis is displayed in a symmetric square root scale.}
\end{figure*}

We first consider the large collection of data including the $1s2p \, ^{1}P_1 \rightarrow 1s^2 \, ^{1}S_0$ ({w}), $ 1s2p \, ^{3}P_2 \rightarrow 1s^2 \, ^{1}S_0$  ({x}), $1s2p \, ^{3}P_1 \rightarrow 1s^2 \, ^{1}S_0$ ({y}), and $1s2p \, ^{3}S_1 \rightarrow 1s^2 \, ^{1}S_0 $ ({z}) radiative transitions in He-like HCI and their theoretical predictions from \cite{asyp2005, Yerokhin2022}. The differences between theory and experiment for each available data point are considered with their associated uncertainties, and the theory-experiment differences are plotted as a function of $Z$ as shown in Fig.~\ref{fig:fit}.  Bayesian analysis is implemented to determine whether the differences between theory and experiment can be reliably modeled by a function of the form $Z^k$, and if so, to obtain the most probable value of $k$. This functional form is chosen because, if a divergence of this form is found, by determining the value of $k$ one could deduce some indication of the kind of missing physics at the origin of the discrepancy, as previously evoked in the literature~\cite{ckgh2012,cpgh2014}\added{, like the presence of an new interaction related to a Yukawa-type potential or milli-charged particles}.

Contrary to previously implemented frequentist approaches, the Bayesian model selection method (see e.g. \cite{vonderLinden,Sivia,Trotta2008,vonToussaint2011}) allows to assign a probability evaluation to each model.
As explained in more detail below, such a probability is evaluated from the integration of the likelihood function, while frequentist approaches only consider the maximum of the likelihood function. 
The latter can be particularly critical when the likelihood function exhibits local maxima.
Moreover, standard frequentist model tests \cite{Neyman1933,Yates1934,Bevington,Massey1951,Akaike1974,Spiegelhalter2002} just apply criteria to choose between pairs of models. 
The quantitative evaluation of model probabilities is particularly helpful for cases that aren't clearly in favor of a particular modeling of the possible data deviation.
Moreover, weighted averages among models become possible within this framework.

The probability of a model $\mathcal{M}$ for a given data set $ \{x_i, y_i\} $ and some background information $I$ is given by the Bayes' theorem 
\begin{equation}
P[\mathcal{M} | \mathrm{Data} ,I] = \frac{P[ \mathrm{Data} | \mathcal{M}, I] \ P[\mathcal{M} | I]}{P[ \mathrm{Data} |  I]}, \label{eq:bayes-data}
\end{equation}
where $P[\mathcal{M} | I]$ is the prior model probability, $P[ \{x_i, y_i\} | \mathcal{M}, I]$ is the model \textit{Bayesian evidence}, commonly called also \textit{marginal likelihood}, and where $P[ \mathrm{Data} |  I]$ is a normalization factor.
The Bayesian evidence can be written as a function of the standard likelihood $P[ \{x_i, y_i\} | \boldsymbol{\theta},\mathcal{M}, I]$ and the prior probability of the model parameters $P[\boldsymbol{\theta}| \mathcal{M}, I]$ with
\begin{multline}
P[ \mathrm{Data} | \mathcal{M}, I]  = \int P[ \mathrm{Data} | \boldsymbol{\theta},\mathcal{M}, I] P[\boldsymbol{\theta}| \mathcal{M}, I] d^{N}\boldsymbol{\theta}. \label{eq:evidence}
\end{multline}
The vector $\boldsymbol{\theta}$ indicates the $N$ parameters of the considered model.

To compare the different deviation models with the null hypothesis (no deviation from QED prediction), we consider the logarithm of the Bayesian evidence ratio $\ln E$ (the Bayes factor) between the two models, here called \textit{Relative Bayesian evidence}. 
The values of relative Bayesian evidence can be related to the more commonly used p-values and standard deviations \cite{Gordon_2007}. 
For practical convenience, in Table \ref{tab:ns-correlation} we report the relation between evidence differences, standard deviations, and p-values from \cite{Gordon_2007}. 
\setlength{\tabcolsep}{12pt}
\begin{table}[tb] 
\begin{tabular}{lrr} 
\hline
\multicolumn{1}{c}{p-value} & \multicolumn{1}{c}{$\log \bar{E}$} & \multicolumn{1}{c}{Standard deviation} \\
\hline
0.04                        & 1.0                                & 2.1                          \\
0.006                       & 2.5                                & 2.7                          \\
0.0003                      & 5.0                                & 3.6                          \\
\hline
\end{tabular}
\caption{
\label{tab:ns-correlation} Correlation between the p-value, Bayesian relative $\log$ evidence, and sigma deviations. Adapted from \cite{Gordon_2007}.}
\end{table}

As anticipated before, we consider a power function model $f_k(Z) = aZ^k$, as used in previous analysis and as expected to describe potential missing QED or electron correlation contributions. In addition, we also consider the same functions plus a constant $f^{c}_k(Z) = aZ^k+b$, which could account for other unexpected missing terms and allows to control the methodology with a different number of parameters. 
As prior parameter probabilities $P[\boldsymbol{\theta}| \mathcal{M}, I]$, we consider a flat distribution that covers the $\pm 5\sigma$ range for $Z=92$, with $\sigma$ equal to the uncertainty of the least accurate measurement available for this $Z$ value. 
Tests on different boundary conditions showed that the final results are not sensitive to the choice of such large priors.
The computation of the multi-dimensional integral in \eqref{eq:evidence} is carried out by the code \textsc{nested\_fit} \cite{trassinelli2017, trassinelli2019, tac2020,Maillard2023} based on the nested sampling algorithm  \cite{Skilling2004,Skilling2006,Ashton2022}.

\section{Results and discussion}

The analysis has been carried out on three different data sets to compare our results with previous works.
The differences between theory and experiment are summarized in Fig.~\ref{fig:fit}.
For the theoretical QED predictions we use values  from Ref.~\cite{Yerokhin2022}  for $5 \leq Z \leq 30$, and from Ref. \cite{asyp2005} for $Z>30$. 
The uncertainty on the energy differences between experimental and theoretical values is obtained via a $L_1$ norm of both uncertainty sources $\delta E_{Total} = \delta E_{Exp} + \delta E_{Theo}$.
To directly compare our Bayesian analysis to past frequentist ones, we first consider the He-like ion {w} transitions measured before 2012, \ie the same set as in Ref. \cite{ckgh2012} (without making any average on measurements on the same elements). We then consider all presently available {w} transitions, and finally all $n=2 \to 1$ transitions presently available.

\begin{figure}[tb]
\includegraphics[width=\linewidth]{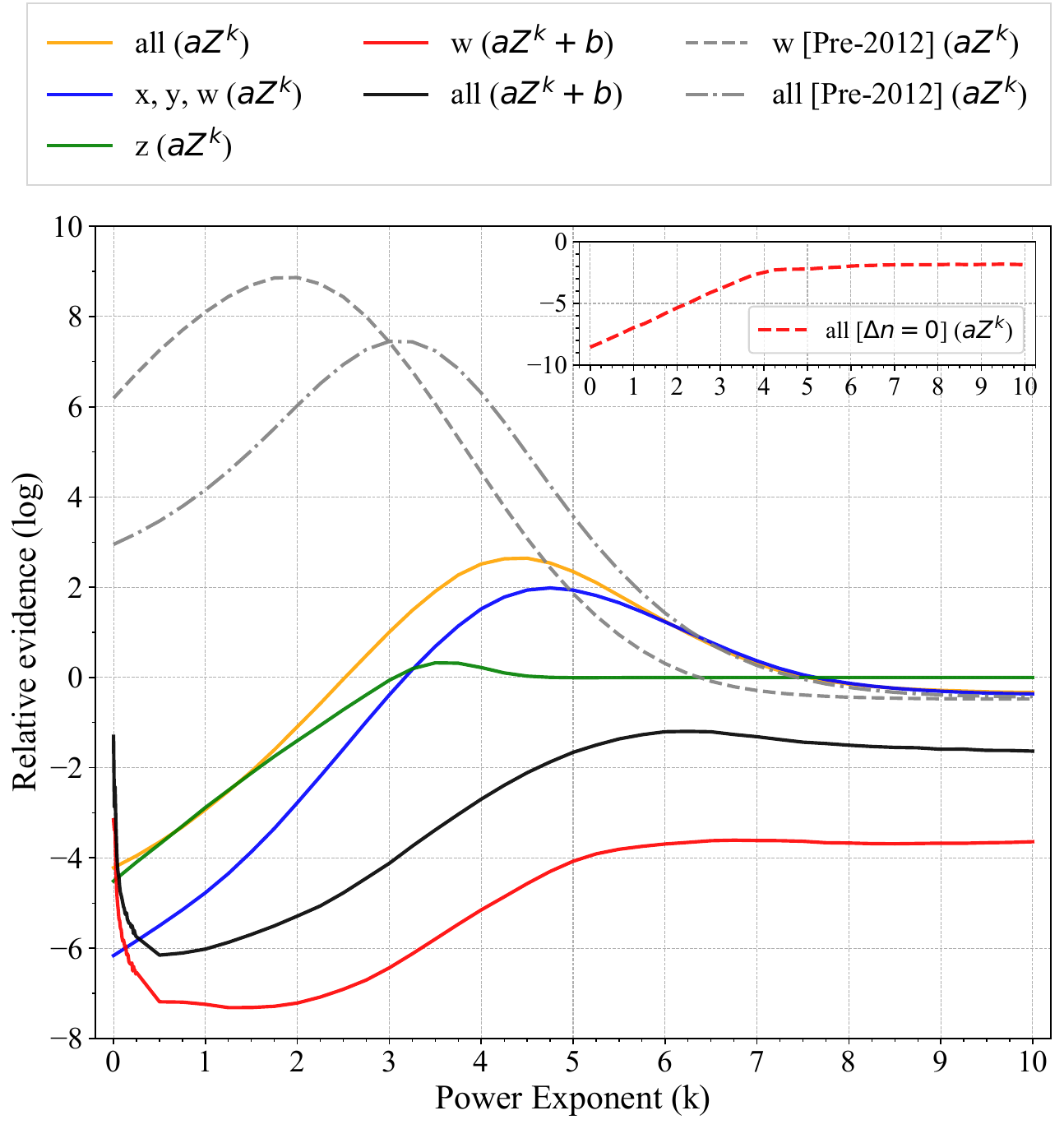}
\caption{\label{fig:comparison}Evidence of multiple test functions on pre-2012 and all datasets for the He-like HCI's ({w} transitions isolated as well). These values are relative (Bayes ratio) to the likelihood test for the QED hypotheses (\ie $\log(E_k) - \log(E_0)$, where $E_0$ is the evidence for the QED constant theory on the measurements). The upper right inset plot shows the same analysis but for the transitions where $\Delta n = 0$ between the initial and final states.}
\end{figure}

The computed relative Bayesian evidence values (in logarithmic scale) are presented in Fig.~\ref{fig:comparison}. Considering first the pre-\num{2012} data used in Ref. \cite{ckgh2012}, one sees a clear preference for a divergence between theory and experiment with a relative logarithmic evidence near $9.0$ for $k=3.5$, when compared to the null hypothesis. This corresponds to a deviation of approximately $4.5\sigma$. This divergence, noted in  Ref.~\cite{ckgh2012}, was taken to be evidence for missing QED contributions in the theory. 
However, when  considering all data, including those gathered since 2012, which corresponds to including \num{25} more values, the significance of this divergence is reduced to $2.7\sigma$ for a value $k=4.5$. 
With a deviation of $2.7\sigma$ there is still some moderate indication of a general trend, but with discrepancies significantly reduced compared to what was previously claimed in the literature. \added{In order to test this finding, we have also computed the relative evidence considering larger error margins for the experimental uncertainty of every value. By adding a factor of \num{2} to the error margins we see no effect in the deviation or in the power exponent.} Due to the S-shell nature of its initial and final states, the {z} transitions were analysed separately. A larger contribution can be seen from {x}, {y} and {w} than from {z}, this suggests there could be a different physical mechanism behind the overall $n = 2 \to 1$ transitions if the discrepancy is to be taken into account. 
The $f^{c}_0(Z)=aZ^{k} + b$ description had a significantly lower Bayesian evidence, with a Bayesian ratio value that corresponds to $4 \sigma$, as may be seen in Fig.~\ref{fig:comparison}. This can be partially understood because a function with three fitting parameters, instead of two, results in a smaller base evidence for the model that describes the data.
\added{Fig. \ref{fig:comparison} shows that the relative evidence (in log) for the fits that include a constant offset is always lower than zero when compared to the baseline theoretical model, indicating that such fits are disfavored by the data.}

The most probable functions $f_k(Z)$ are represented in Fig.~\ref{fig:comparison} together with the differences between theory and experiment.
The median value of the scaling factor $a$ is plotted together with the corresponding 99\% confidence region. The mean value of the parameter cannot be used as this assumes that the data are normally distributed, which is not generally the case especially for small datasets.  One can see that with the addition of higher precision measurements post-2012, the best-fit model order, $k$, went up and the linear coefficient, $a$, went down, which indicates that the discrepancy is more in-line with the current theoretical state of the art. The model's linear coefficient is mostly influenced by the precision of the high-Z measurements, as these were the main limitation to the deviation prediction. The evidence of the model is highly dependent on high-Z data and their experimental uncertainty.

A similar analysis was also performed for the $\Delta n = 0$ transitions, with a total of \num{116} data points.
For this case, no particular indication of a deviation is found, as the relative evidence (in log) was always lower than zero considering the theory baseline.
This is in accordance with the current QED being that only $\ell = 0$ states are involved. The results of the evidence logarithm are shown in the upper right inset of Fig.~\ref{fig:comparison}. 
No deviation from QED theory are visible for such transitions.

\begin{figure}[tb]
\includegraphics[width=\linewidth]{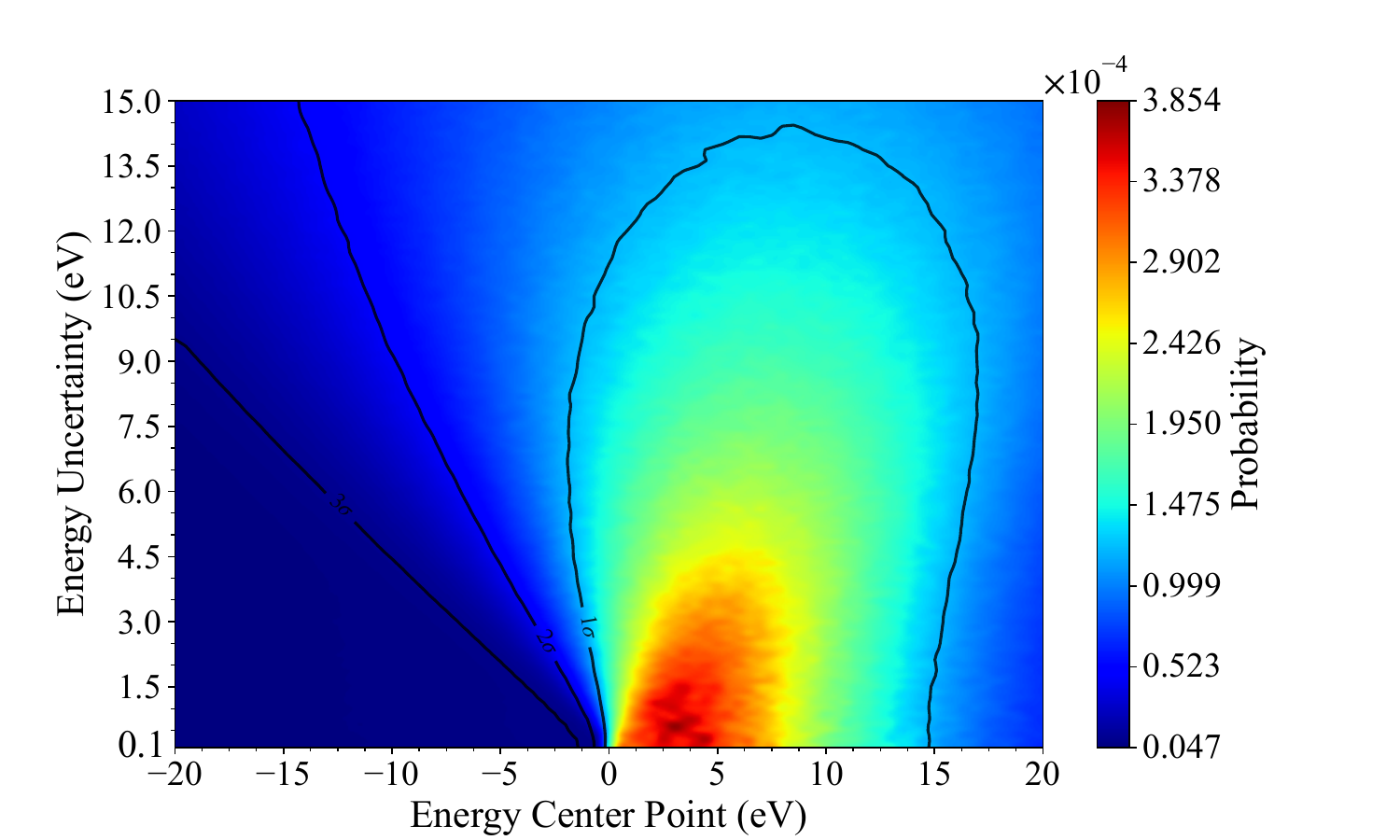}
\includegraphics[width=\linewidth]{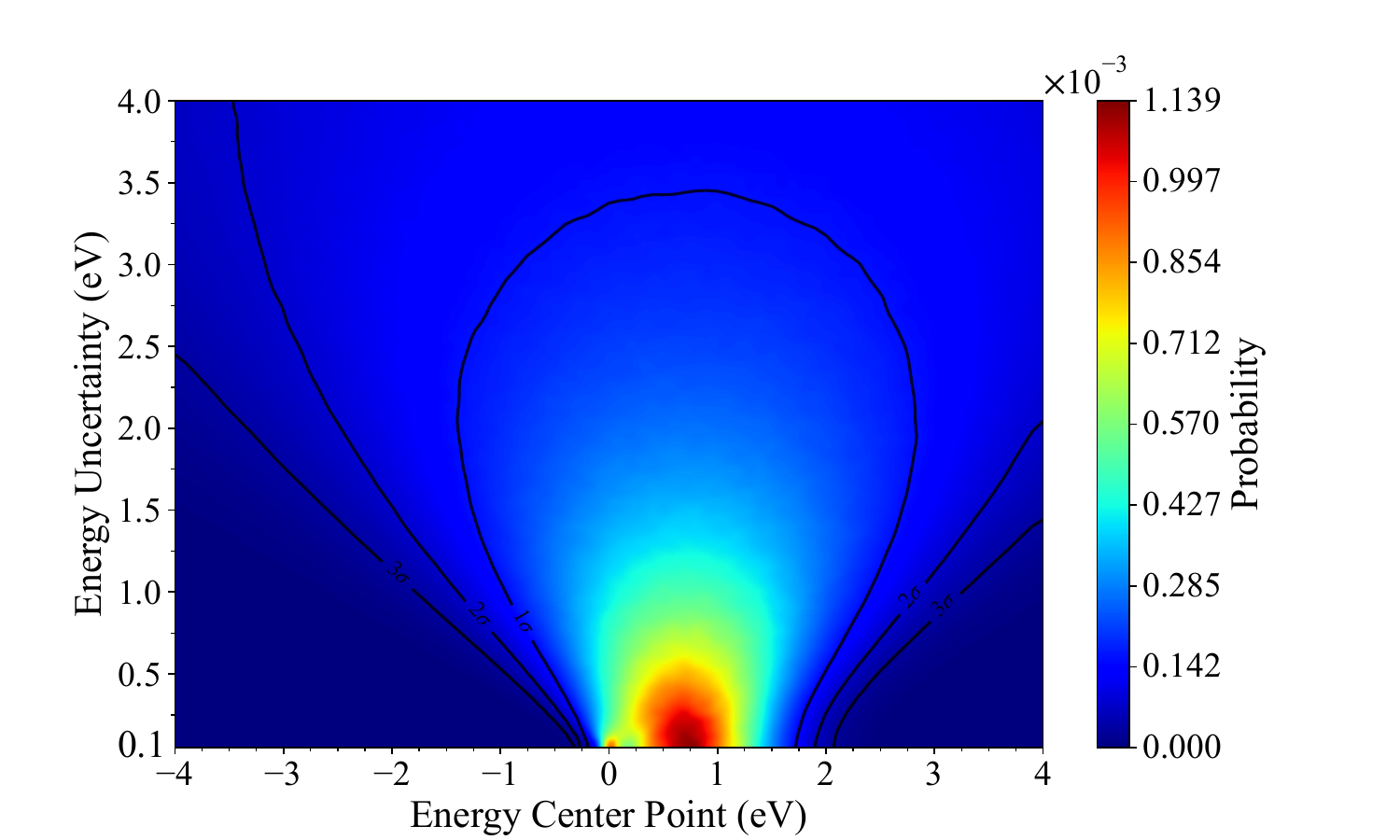}

\caption{\label{fig:mean-sig-Xe}Probability of a new U$^{90+}$  (up) and Xe$^{52+}$ (bottom) hypothetical HCI measurement value.
New points closer to the red region favor a power tendency, while the other ones mean that the value is closer to the current theory (\cite{asyp2005, Yerokhin2022}) for an average of all power exponents ($k$).
}
\end{figure}

Having determined that there is some, albeit much reduced, evidence for a theory-experiment discrepancy that behaves as $Z^{k}$, following an approach similar to the one discussed in \cite{vonToussaint2011}, we can use the same methodology backwards to determine which new measurements should be performed and with which minimal accuracy to have the largest impact in constraining this discrepancy.
For this evaluation, we chose as possible measurements the ions U$^{90+}$, Pb$^{80+}$ and Xe$^{52+}$, knowing that in principle the high-$Z$ species should have the greatest impact due to the $Z^{n}$ scaling, and because of the lack of measurements for high $Z$.
To calculate such a probability, we consider the same data sets as used above, plus an additional hypothetical datum value $E\pm\Delta E$ for an ion with a chosen $Z$. 
This datum is used as a probe point to gather information about the evidence if there were to be a measurement with those values in the data.
With $Z$ fixed, the fictive measurement parameters $E$ and $\Delta E$ are swept across a range of possible values to obtain evidences of the $f_k(Z)$ for each value of $k$. 
The final probability is obtained by averaging between the different models, \ie over the different values of $k$.

The probability that results from a hypothetical new measurement of a transition in  U$^{90+}$ and  Xe$^{52+}$ is shown in Fig.~\ref{fig:mean-sig-Xe}.
The isoprobability contours for \num{1}, \num{2} and \num{3} standard deviations from the average are also shown. The details of the formulas are presented in the Appendix. 
The figures show a clear asymmetry of the probability distribution along zero, which corresponds to the direction of the possible deviation from the null hypothesis, while still remaining compatible with it.
These figures show also that future experiments should provide a sufficiently small uncertainty to confirm or not the present measurement trend. 
In other words, to be relevant for future tests, new experiments should have an uncertainty small enough to discriminate between the different regions visible in the figures. 
In the case of uranium, this means that future measurements should have an accuracy better than 10~eV.
In the case of xenon, such an accuracy should be reduced to 1~eV or lower.
For the case of lead, not shown in the figures, the accuracy should be around 5~eV.

\section{Conclusions}

We have re-evaluated the current status of the agreement between theory and experiment for transitions in He-like ions using a Bayesian approach. While we confirm that the data up to \num{2012} suggested a large, significant discrepancy between theory and experiment with a dependence that goes as $Z^{k}$, results in the last decade have significantly reduced this discrepancy. No deviation from QED prediction can currently be claimed. Moreover, the present study indicates that medium and high-$Z$ measurements are strongly required to support the evidence for any possible deviation.

On the other hand, in light of past measurements, for future experimental design, one can use the predictions of the probability of future hypothetical measurements to
determine if their accuracy will be sufficient to make the experiment relevant for finding a strong evidence for a deviation with theoretical predictions. 

A potential future analysis would involve looking deeper into the atomic structure and classifying these shifts based on the types of atomic orbitals, for example into all S and P states.

\section*{Acknowledgments}
N. P. and P. I thanks the CNRS Institute of Physics for support. P. I. is a member of the Allianz Program of the Helmholtz Association, Contract No. EMMI HA-216“Extremes of Density and Temperature: Cosmic Matter in the Laboratory”. Laboratoire Kastler Brossel is a Unité Mixte de Recherche n° 8552 involving Sorbonne Université, CNRS, ENS-PSL Research University and Collège de France.

We acknowledge support from the PESSOA Huber Curien Program 2022, Number 47863UE. This research was funded in part by Fundação para a Ciência e Tecnologia (FCT) (Portugal) through the research center grant UID/FIS/04559/2020 to LIBPhys-UNL from the FCT/MCTES/PIDDAC, Portugal. C.G. acknowledges support from FCT, under Contracts No. 2022.11197.BD

\appendix*

\section{Probability formulas}
From a set of given data and possible models, what is the probability $P[ (x,y,\sigma) | \mathrm{Data}]$ for having a new experimental point $(x,y)$ with a defined output and accuracy $\sigma$ independently from the most adapted model? 
Considering a set of $M$ reasonable models $\{\mathcal{M}_m\}$, $P[ (x,y,\sigma) | \mathrm{Data}]$ can be written as
\begin{multline}
P[ (x,y,\sigma) | \mathrm{Data} ]   = 
\sum_{m=1}^M P[ (x,y,\sigma), \mathcal{M}_m | \mathrm{Data} ] = \\
\cfrac{\sum_{m=1}^M P[ (x,y,\sigma), \mathrm{Data} |\mathcal{M}_m ] \,  P[\mathcal{M}_m]}{P[\mathrm{Data}]}.
\end{multline}

$P[ (x,y,\sigma), \mathrm{Data} |\mathcal{M}_m ]$ is the evidence relative to the data including the new hypothetical datum $(x,y,\sigma)$.
The normalization factor can be found imposing 
\begin{equation}
    \int P[ (x,y,\sigma) | \mathrm{Data} ] \mathrm{d} y \, \mathrm{d} \sigma = 1
\end{equation}
Assuming a uniform model probability $P[\mathcal{M}_m]=1/M$, we have
\begin{equation}
   P[\mathrm{Data}] =
   \cfrac 1 M \sum_{m=1}^M  
   \int_{y_\mathrm{min}}^{y_\mathrm{max}} \hspace{-0.6cm}  \mathrm{d} y 
   \int_0^{\sigma_\mathrm{max}} \hspace{-0.6cm} \mathrm{d} \sigma \,
   P[ (x,y,\sigma), \mathrm{Data} |\mathcal{M}_m ],
\end{equation}
which can easily be transformed into a discrete form.

\newpage

\bibliographystyle{apsrev}

\bibliography{bib}
\end{document}